\documentclass[aps,prd,onecolumn,groupedaddress,showpacs,nofootinbib,amssymb]{revtex4}
\usepackage[dvips]{graphicx}
\usepackage{amssymb}
\usepackage{amsmath}
\usepackage{graphicx}
\usepackage{amsfonts}
\usepackage{bm}

\begin{document}

\title{Singular Bouncing Cosmology from Gauss-Bonnet Modified Gravity}

\author{V.K. Oikonomou$^{1,2}$}\,\thanks{v.k.oikonomou1979@gmail.com}
\affiliation{
$^{1)}$ Tomsk State Pedagogical University, 634061 Tomsk \\
$^{2)}$  Lab. Theor. Cosmology, Tomsk State University of Control Systems\\ 
and Radioelectronics, 634050 Tomsk, Russia (TUSUR)
}

\begin{abstract}
We study how a cosmological bounce with a Type IV singularity at the bouncing point, can be generated by a classical vacuum $F(G)$ gravity. We focus our investigation on the behavior of the vacuum $F(G)$ theory near the Type IV singular bouncing point and also we address the stability of the resulting solution, by treating the equations of motion as a dynamical system. In addition, we investigate how the scalar perturbations of the background metric evolve, emphasizing to cosmological times near the Type IV singular bouncing point. Finally, we also investigate which mimetic vacuum $F(G)$ gravity can describe the singular bounce cosmology.
\end{abstract}


\maketitle

\section{Introduction}

The observational data that came into play the last twenty years, have made compelling the consistent description with a theory that can harbor all the different observationally verified phenomena. Two are the main observations that need to be described theoretically, the late-time acceleration verified in the late 90's \cite{riess}, and the early-time acceleration. With regards to the latter, there is much way to be covered until we conclude whether inflation ever existed. However, the latest Planck data \cite{planck} pose severe constraints on inflationary models, and indicate which features should a consistent theory of inflation have. Modified gravity theories offer a promising and solid theoretical framework that can consistently describe late-time and early-time acceleration, for reviews on this vast issue see \cite{reviews1}. Among the most promising are the $F(R)$ theories of gravity, which also offer the possibility to describe simultaneously early and late-time acceleration \cite{sergei2003}. The $F(R)$ gravity is the simplest modification of Einstein-Hilbert gravity, since instead of having simply the Ricci scalar $R$ in the Lagrangian, a function of $R$ appears. In four dimensions, instead of this simple modification of Einstein-Hilbert gravity, there also exist other theoretical descriptions that are also promising, like the $F(G)$ theories of gravity, with $G$ being the Gauss-Bonnet invariant $G=R^2-4R_{\mu \nu}R^{\mu \nu}+R_{\mu \nu \rho \sigma}R^{\mu \nu \rho \sigma}$, and with $R_{\mu \nu}$, $R_{\mu \nu \rho \sigma}$ standing for the Ricci tensor and the Riemann tensor respectively. Although in principle the resulting equations of motion are expected to have fourth order derivatives of the metric tensor, it turns out that the theory contains only second order derivatives, and therefore it is rendered not too complicated to be studied. Particularly, it has been shown that within the context of $F(G)$ theory, late-time acceleration can be achieved \cite{sergegauss1,sergegauss2,sergegauss3,sergegauss4,fg4,fg5,fg6,fg7,fg8}. For informative reviews on $F(G)$ theory of gravity see \cite{reviews1}.

On the other hand, an appealing alternative to the standard inflationary paradigm, comes from the bounce cosmology theories \cite{bounce1,bounce2,bounce3,bounce4,bounce5,quintombounce,ekpyr1,superbounce2,matterbounce}. In this kind of theories, one of the most severe drawbacks of the inflationary paradigm, the initial singularity problem, is absent, and it is conceivable that this feature makes them quite appealing. The initial singularity is a crushing type timelike  singularity, and at the point it occurs the geodesics cannot be continuously be extended, that is, geodesics incompleteness occurs. The singularity theorems of Hawking and Penrose \cite{hawkingpenrose} fully describe these singularities, and much work has been devoted in studying spacelike singularities. In cosmology however, most of the singularities are timelike, and the most severe of these are the initial singularity and the Big Rip \cite{ref5}. With regards to the latter, it is the most severe of the four types of finite singularities classified in \cite{Nojiri:2005sx}. Also sudden singularities were also studied in \cite{barrowsing1,barrowsing2,barrowsing3,barrow}. Among the four types of finite-time singularities, the most mild from a phenomenological point of view, is the Type IV singularity, at which no geodesic incompleteness occurs. For recent studies on this kind of singularity see \cite{Barrow:2015ora,noo1,noo2,noo3,noo4,noo5}. The most phenomenologically interesting feature of theories with Type IV finite time singularities is that the Universe may smoothly pass through \cite{noo4,noo5} these timelike singularities, without any catastrophic consequences. In some cases, there might be observational evidence or indication of the passage of the Universe through a Type IV singularity \cite{noo4}, but it is important that this is not catastrophic and it always is a smooth  passage \cite{noo5}.

In view of the aforementioned interesting features of the bouncing cosmologies, in this paper we aim to study a specific bounce which contains a Type IV singularity at the bouncing point. In this way, the cosmological system is free from the initial singularity, but at the same time we investigate the implications of such a mild finite time singularity. Particularly, we shall investigate how such a cosmology can be described by a classical vacuum $F(G)$ gravity, with special emphasis being given on how the $F(G)$ gravity behaves near the Type IV singular bouncing point\footnote{From this point, when we refer to the bouncing point or the singular point, we refer to the same point.}. Also, we investigate how the resulting solution behaves, by checking the stability of the resulting equations, when these are viewed as a dynamical system. In this case we examine if the solution is the final attractor of the system. It is conceivable of course that since the Universe passed through the singular point smoothly, the resulting $F(G)$ gravity should not be the final attractor of the theory, so in some sense instability of the dynamical system is anticipated. In addition, we shall investigate how the scalar perturbations of the background metric behave for the case of the resulting vacuum $F(G)$, near the Type IV singularity. Finally, we shall investigate which vacuum $F(G)$ gravity generates the same singular bounce we are discussing, but in the context of mimetic $F(G)$ theory, which was developed in \cite{sergeioikonomoumimetic}. Again the focus will be for cosmic times near the singular bouncing point. We believe that our work will provide some new information on the behavior of classical modified theories of gravity near a Type IV singular bouncing point, a study which combines bouncing cosmology with a mild singularity at the bounce, and also a classical description with an $F(G)$ gravity.

This paper is organized as follows: In section II we briefly review the basic features of a Type IV singular bounce cosmology, and in section III by using known cosmological reconstruction techniques, we investigate which vacuum $F(G)$ can describe the Type IV singular bounce, by focusing on the behavior near the bouncing point. Also, we discuss the possible connection of the resulting $F(G)$ theory with other viable $F(G)$ theories. In section IV we study the stability of the resulting solution we found in section III, by treating the system of equations of motion as a dynamical system. To this end, we rewrite the system of equations of motion in terms of new variables, which make the study more clear from a physical point of view. In section V we investigate how the scalar perturbations of the background metric behave, for the vacuum $F(G)$ theory we found in section III, emphasizing again on the behavior near the Type IV singular point. Finally, in section VI we investigate which mimetic $F(G)$ gravity can describe the Type IV singular bounce cosmology, focusing again on the behavior near the Type IV singularity. The conclusions along with a discussion follow in the end of the paper.

\section{A Brief Description of the Singular Bounce}

As we already mentioned, bouncing cosmology \cite{bounce1,bounce2,bounce3,bounce4,bounce5,quintombounce,ekpyr1,superbounce2,matterbounce} is an appealing alternative scenario to the standard inflationary paradigm, with the most attractive feature of bouncing cosmology being the fact that there is no initial singularity, and there exists also the possibility of successfully describing early-time acceleration \cite{matterbounce,ekpyr1}. It has been recently shown however, that other types of milder singularities \cite{Barrow:2015ora,noo1,noo2,noo3,noo1,noo5} might occur during the cosmological evolution, without having the catastrophic consequences of the crushing type singularities, like the Big Rip one \cite{ref5}. Particularly, in Ref. \cite{noo5} it has been demonstrated that a Type IV singularity \cite{Nojiri:2005sx,noo1,noo2,noo3,noo4,noo5} may occur at the bounce point of a general bouncing cosmology, having interesting consequences. We shall briefly describe this possibility in this section, in order to render the presentation self-contained.

Firstly, a cosmological bounce can be separated in two evolution eras, the contraction era, during which the scale factor decreases ($\dot{a}<0$), and the expansion era, during which the scale factor increases ($\dot{a}>0$). In between the two eras, and after the contraction era, the Universe reaches a minimal radius, where $\dot{a}=0$, and this is the reason that the bouncing scenario is free of the initial singularity. In principle, the bouncing point, that is, the point at which the bounce occurs, can freely be chosen, so we assume that the bounce occurs at $t=t_s$, so when $t<t_s$, the Hubble rate $H=\dot{a}/a$ is negative (since $\dot{a}<0$), while for $t>t_s$ it is positive (since $\dot{a}>0$), and of course at the bounce it becomes equal to zero $H(t_s)=0$. We shall consider the bouncing cosmology with scale factor,
\begin{equation}\label{scalebounce}
a(t)=e^{f_0\left(t-t_s\right)^{2(1+\varepsilon)}},
\end{equation}
with $a(t_s)=1$, and $f_0$ an arbitrary parameter. In addition, the parameter $\varepsilon$ is assumed to be $\varepsilon < 1$, so the bouncing cosmology of Eq. (\ref{scalebounce}) is assumed to be a deformation of the well known symmetric bounce \cite{sergeibabmba},
\begin{equation}\label{bambabounce}
a(t)=e^{f_0\left(t-t_s\right)^{2}}.
\end{equation}
From Eq. (\ref{scalebounce}) it easily follows that the Hubble rate is equal to,
\begin{equation}\label{hubblebounce}
H(t)=2 (1+\varepsilon ) f_0 \left(t-t_s\right)^{2\varepsilon+1 }.
\end{equation}
For simplicity we introduce the following variables,
\begin{equation}\label{identifica}
\beta=2 (1+\varepsilon ) f_0,\,\,\,\alpha=2\varepsilon+1\, ,
\end{equation} 
so the Hubble rate becomes, 
\begin{equation}\label{hubratepresentpaper}
H(t)=\beta \left(t-t_s\right)^{\alpha }\, .
\end{equation}
According to the classification of finite time singularities, the following types of singularities occur at $t=t_s$ which is the bouncing point,
\begin{itemize}\label{lista}
\item $\alpha<-1$ corresponds to the Type I singularity.
\item $-1<\alpha<0$ corresponds to Type III singularity.
\item $0<\alpha<1$ corresponds to Type II singularity.
\item $\alpha>1$ corresponds to Type IV singularity.
\end{itemize}
so the relevant to us case is when $\alpha>1$, and since we want to render the bounce of Eq. (\ref{scalebounce}) a deformation of the symmetric bounce (\ref{bambabounce}), we further assume that $1<\alpha<2$ (or equivalently $0<\varepsilon< \frac{1}{2}$), something which implies that the second derivative of the Hubble rate (\ref{hubratepresentpaper}) diverges. As shown in \cite{noo4}, this can have interesting phenomenological consequences. Finally, we need to make sure that the Hubble rate and the scale factor never become complex, so the parameter $\alpha$ needs to be appropriately chosen, so we make the choice $\alpha=13/11$, but in general $\alpha=\frac{2n-1}{2m+1}$. For a detailed account on that see \cite{noo5}.

\section{Singular Bounce from $F(G)$ Gravity}

Having described the cosmological bounce with a Type IV singularity at the bouncing point, in this section we shall investigate how this type of cosmological evolution can be described in terms of a vacuum $F(G)$ gravity. Special emphasis shall be given in the $F(G)$ gravity that describes the bounce near the Type IV singularity, since the general problem is rather difficult to address, due to lack of analytic solutions of the corresponding differential equations.

In order to find the $F(G)$ gravity that describes the bounce near the singular point, we shall use some very well know reconstruction techniques for $F(G)$ theories of gravity \cite{fg4,fg5,fg6}. For a similar technique to the one we shall use here, see \cite{fg7}. The Jordan frame $F(G)$ gravity action in the absence of matter fluids (vacuum $F(G)$), is equal to \cite{reviews1,fg4,fg5,fg6}, 
\begin{equation}\label{actionfggeneral}
\mathcal{S}=\frac{1}{2\kappa^2}\int \mathrm{d}^4x\sqrt{-g}\left ( R+F(G)\right )\, ,
\end{equation}
where $\kappa^2=1/M_{pl}^2$, with $M_{pl}=1.22\times 10^{19}$GeV. By varying with respect to the metric, the gravitational equations read,
\begin{align}\label{fgr1}
& R_{\mu \nu}-\frac{1}{2}g_{\mu \nu}F(G)-\Big{(}-2RR_{\mu \nu}+4R_{\mu \rho}R_{\nu}^{\rho}-2R_{\mu}^{\rho \sigma \tau}R_{\nu \rho \sigma \tau}+4g^{\alpha \rho}g^{\beta \sigma}R_{\mu \alpha \nu \beta}R_{\rho \sigma}\Big{)}F'(G)\\ \notag &
-2 \left (\nabla_{\mu}\nabla_{\nu}F'(G)\right )R+2g_{\mu \nu}\left (\square F'(G) \right )R-4 \left (\square F'(G) \right )R_{\mu \nu }+4 \left (\nabla_{\mu}\nabla_{\nu}F'(G)\right )R_{\nu}^{\rho }\\ \notag &+4 \left (\nabla_{\rho}\nabla_{\nu}F'(G)\right ) R_{\mu}^{\rho}
-4g_{\mu \nu} \left (\nabla_{\rho}\nabla_{\sigma }F'(G)\right )R^{\rho \sigma }+4 \left (\nabla_{\rho}\nabla_{\sigma }F'(G)\right )g^{\alpha \rho}g^{\beta \sigma }R_{\mu \alpha \nu \beta }=0
\end{align}
where the Gauss-Bonnet invariant expressed in terms of the Hubble rate, equals to, 
\begin{equation}\label{gausbonehub}
G=24H^2\left (\dot{H}+H^2 \right )\, .
\end{equation}
Assuming a flat Friedmann-Robertson-Walker (FRW) background, with line element, 
\begin{equation}\label{metricformfrwhjkh}
\mathrm{d}s^2=-\mathrm{d}t^2+a^2(t)\sum_i\mathrm{d}x_i^2\, ,
\end{equation}
the gravitational equations of Eq. (\ref{fgr1}) take the following form,
\begin{align}\label{eqnsfggrav}
& 6H^2+F(G)-GF'(G)+24H^3\dot{G}F''(G)=0\\ \notag &
4\dot{H}+6H^2+F(G)-GF'(G)+16H\dot{G}\left ( \dot{H}+H^2\right ) F''(G)
\\ \notag & +8H^2\ddot{G}F''(G)+8H^2\dot{G}^2F'''(G)=0\, .
\end{align}
The reconstruction technique presented in \cite{fg4,fg6}, employs the use of an auxiliary field denoted as $\phi$, which as was shown can be identified with the cosmic time $t$, that is $\phi=t$. By introducing two proper functions of $t$, namely $P(t)$ and $Q(t)$, the Jordan frame action of Eq. (\ref{actionfggeneral}) becomes,
\begin{align}\label{actionfrg}
& \mathcal{S}=\frac{1}{2\kappa^2}\int \mathrm{d}^4x\sqrt{-g}\left ( R+P(t)G+Q(t)\right )\, ,
\end{align}
and by varying it with respect to $t$, we obtain,
\begin{equation}\label{auxeqnsvoithitiko}
\frac{\mathrm{d}P(t)}{\mathrm{d}t}G+\frac{\mathrm{d}Q(t)}{\mathrm{d}t}=0\, .
\end{equation}
Having Eq. (\ref{auxeqnsvoithitiko}) at hand, solving it with respect to $t=t(G)$ and by substituting this in the following expression, 
\begin{equation}\label{ebasc}
F(G)=P(t)G+Q(t)\, ,
\end{equation}
we finally have the $F(G)$ gravity. It is therefore obvious that the functions $P(t)$ and $Q(t)$ play an important role in the determination of the $F(G)$ gravity, so we now derive the differential equations that yield these functions. By combining Eq. (\ref{ebasc}) and the first of the equations appearing in Eq. (\ref{eqnsfggrav}), we obtain the following differential equation,
\begin{align}\label{ak}
& Q(t)=-6H^2(t)-24H^3(t)\frac{\mathrm{d}P}{\mathrm{d}t}\, ,
\end{align}
and finally combining Eqs. (\ref{ebasc}) and (\ref{ak}), we obtain the following differential equation,
\begin{align}\label{diffept}
& 2H^2(t)\frac{\mathrm{d}^2P}{\mathrm{d}t^2}+2H(t)\left (2\dot{H}(t)-H^2(t) \right )\frac{\mathrm{d}P}{\mathrm{d}t}+\dot{H}(t)=0\, .
\end{align}
When solved, Eq. (\ref{diffept}) determines the function $P(t)$, and therefore $Q(t)$, so upon substitution of $P(t)$ in Eq. (\ref{ak}), we get $Q(t)$. Finally, from Eq. (\ref{auxeqnsvoithitiko}) we get the function $t=t(G)$, and by substituting that in Eq.(\ref{ebasc}) we obtain the final form of the $F(G)$ gravity. We now apply this method in order to find the $F(G)$ gravity which describes the bounce near the bouncing point.

\subsection{$F(G)$ Gravity Near the Bouncing Point}

The general problem of finding the $F(G)$ gravity for the singular bounce with Hubble rate (\ref{hubratepresentpaper}), is rather difficult to deal with analytically, so we shall focus on finding the $F(G)$ gravity near the bouncing point $t=t_s$, which recall that is the point where the Type IV singularity occurs too. Consequently, at some point, we shall specify our analysis around the singularity.

For the Hubble rate of Eq. (\ref{hubratepresentpaper}), the differential equation (\ref{diffept}) that yields the function $P(t)$ can be shown that it becomes,
\begin{align}\label{diffept2}
& \frac{2  \beta }{\alpha }(t-t_s)^{1+\alpha }\frac{\mathrm{d}^2P}{\mathrm{d}t^2}+4 (t-t_s)^{\alpha } \beta+1=0\, ,
\end{align}
which can be solved analytically to yield,
\begin{equation}\label{dfesolu}
P(t)=-\frac{(t-t_s)^{1-2 \alpha } \left((t-t_s)^{\alpha }-2 (t-t_s)^{\alpha } \alpha -2 \beta  C_1+2 \alpha  \beta  C_1\right)}{2 (-1+\alpha ) (-1+2 \alpha ) \beta }+C_2\, ,
\end{equation}
so by substituting Eq. (\ref{dfesolu}) in Eq. (\ref{ak}), we obtain the function $Q(t)$ which appears in Appendix A, since it is too complicated to quote it here. By using the resulting expressions for the functions $Q(t)$ and $P(t)$, the final form of Eq. (\ref{auxeqnsvoithitiko}) becomes, 
\begin{equation}\label{finaformofauxiliaryeq}
\frac{x^{-1-2 \alpha } \left(4 x^{3 \alpha } \alpha  \beta ^3 \left(11 x^{\alpha }-12 C_1 \beta \right)-G x \left(x^{\alpha }-2 C_1 \beta \right)\right)}{2 \beta }=0\, ,
\end{equation}
where we have set $x=t-t_s$ for simplicity. It is obvious that the above equation is rather difficult to be solved analytically, so we shall simplify it by keeping the dominant terms in the limit $x\rightarrow 0$, which corresponds to the limit $t\rightarrow t_s$. Therefore, by taking the limit $x\rightarrow 0$, then Eq. (\ref{finaformofauxiliaryeq}) becomes,
\begin{equation}\label{dgfshsyyss}
C_1 G x^{-2 \alpha }-24 C_1 x^{-1+\alpha } \alpha  \beta ^3=0\, ,
\end{equation}
which yields,
\begin{equation}\label{onyfe}
x=\frac{G^{\frac{1}{(3\alpha -1)}}}{\left(24 \alpha  \beta ^3\right)^{\frac{1}{(3\alpha -1)}}}\, ,
\end{equation}
Then by substituting (\ref{onyfe}) in $P(t)$ and $Q(t)$, and by using Eq. (\ref{ebasc}), we obtain the resulting expression for the $F(G)$ gravity near the singular point, which is,
\begin{equation}\label{actaulfg}
F(G)=C_2G+A G^{\frac{2 \alpha }{-1+3 \alpha }} +B G^{\frac{\alpha }{-1+3 \alpha }}\, ,
\end{equation} 
where the coefficients $A$ and $B$ are given in the Appendix A. We can give however a simpler form by exploiting the fact that we are interested in the limit $t\rightarrow t_s$. The Gauss-Bonnet invariant of Eq. (\ref{gausbonehub}), is written in terms of the variable $x=t-t_s$ we introduced earlier, as follows,
\begin{equation}\label{argausbbbe}
G=24 x^{-1+3 \alpha } \alpha  \beta ^3+24 x^{4 \alpha } \beta ^4\, ,
\end{equation}
from which it is obvious that as $x\rightarrow 0$ (or equivalently $t\rightarrow t_s$), the Gauss-Bonnet invariant tends also to zero. Hence, by keeping the most dominant terms from the $F(G)$ gravity of Eq. (\ref{actaulfg}), we get the small $G$ limit of it,
\begin{equation}\label{fgsmalglimit}
F(G)\simeq C_2 G+B G^{\frac{\alpha }{-1+3 \alpha }}\, .
\end{equation}
Therefore, in the small $G$ limit, or equivalently near the Type IV singularity, which we chose to be the bouncing point, the $F(G)$ gravity that can generate the Hubble rate (\ref{hubratepresentpaper}) is approximately described by the expression of Eq. (\ref{fgsmalglimit}). Since we are discussing for cosmological times near the bouncing point, it is worth examining how the evolution of perturbations behave for this $F(G)$ model. This will be done in detail in a later section.

\subsection{Connection with Other Viable $F(G)$ Gravities and Possible Late-time Behavior}

In the previous section we investigated which $F(G)$ gravity theory can successfully describe the singular bounce cosmology of Eq. (\ref{hubratepresentpaper}), near the singular bounce. However the complexity of the resulting differential equations forced us to find an approximate solution, with the final $F(G)$ being that of Eq. (\ref{fgsmalglimit}), or can be further simplified neat the bouncing point, as 
\begin{equation}\label{gupro}
F(G)\simeq B G^{\frac{\alpha }{-1+3 \alpha }}\, .
\end{equation}
Now, it would be interesting to ask how the complete classical $F(G)$ gravity would look like. A possible answer is that the complete classical $F(G)$ gravity description would be one of the four possible forms of $F(G)$ gravity that lead to finite-time singularities, first studied in \cite{fg5} and further investigated in \cite{lobo,myrzafgfinite}. Particularly, the $F(G)$ gravities that lead to singularities are of the following form \cite{fg5,lobo,myrzafgfinite},

\begin{equation}\label{cand1}
F(G)=\frac{a_1G^n+b_1}{a_2G^n+b_2}\, ,
\end{equation}
\begin{equation}\label{cand2}
F(G)=\frac{a_1G^{n+N}+b_1}{a_2G^n+b_2}\, ,
\end{equation}
\begin{equation}\label{cand3}
F(G)=a_3G^n (b_3G^m+1)\, ,
\end{equation}
\begin{equation}\label{cand4}
F(G)=G^m\frac{a_1G^n+b_1}{a_2G^n+b_2}\, ,
\end{equation}
where the parameters $a_i$, and $b_i$ with $i=1,2,3$, are arbitrary real constants. Obviously, the $F(G)$ gravity we found, which for small $G$ is given by Eq. (\ref{gupro}), can be the limiting case of the above $F(G)$ gravities, which lead to finite singularities for some values of the parameters. For example the $F(G)$ gravity of Eq. (\ref{cand4}) in the small $G$ limit behaves as $\sim \frac{b_1}{b_2}G^m$, which is clearly similar to the one we obtained in Eq. (\ref{gupro}). Note that in the large $G$ limit, the $F(G)$ gravity (\ref{cand4}) behaves $\sim \sim \frac{a_1}{a_2}G^m$, so the late-time behavior of this $F(G)$ gravity is described by a power-law $F(G)$ function and as it was shown in \cite{sergegauss1,sergegauss2,sergegauss3,sergegauss4}, such power-law modified gravity theories can serve as models for dark energy.

Before we close this section, we need to stress that the absence of an analytic solution in the case of a singular bounce is exactly due to the existence of the singularity. In other bouncing cosmologies, for which no singularity occurs, this lack of analyticity no longer persists as a problem, and the exact behavior of the $F(G)$ gravity can be found, like for example in Ref. \cite{sergeibabmba}.

\section{Stability of the $F(G)$ Gravity Solutions Near the Bouncing Point}

The FRW equations of Eq. (\ref{eqnsfggrav}) for the $F(G)$ gravity constitute a dynamical system which determines the behavior of solutions which satisfy these equations. Stability of a solution of this dynamical system against linear perturbations would mean that this solution is one of the final attractors of the theory. On the contrary if a solution of the system is unstable against linear perturbations, then it is obvious that this solution is not a final attractor of the theory. The focus in this section is on the stability of the solutions we found for the $F(G)$, near the bouncing point, against linear perturbations of the solutions. What it is expected is obviously that the solutions we found near the bouncing point are unstable, since the cosmological evolution does not stop at the bouncing point, but continues and the Universe starts to expand. Therefore, the local solutions we found that describe the $F(G)$ gravity near the bouncing point, should be unstable against linear perturbations, since they do not describe the whole evolution, but a small part of it, near the Type IV singularity. In the rest of this section, we shall use some convenient variables and we shall study the stability of the dynamical system of Eq. (\ref{eqnsfggrav}) against linear perturbations.

We adopt the techniques and notation used in \cite{fg7}, so we consider linear perturbations of the solution $g(N)=H^2$, of the following form,
\begin{equation}\label{pert2}
g(N)\rightarrow g(N)+\delta g(N)
\end{equation}
with the function $g(N)$ satisfying the FRW equations (\ref{eqnsfggrav}). Expressing the dynamical system of Eq. (\ref{eqnsfggrav}) in terms of the function $g(N)$, we get,
\begin{align}\label{sfrw12}
& 288g^2(N)F''(G)\Big{[}\left ((g'(N))^2+g(N)\right )g''(N)+4g(N)g'(N)+4g(N)g'(N)\Big{]}\\ \notag &
6g(N)+F(G)-12g(N)\left (g'(N)+2g(N) \right )F'(G)=0\, .
\end{align} 
The conditions that ensure the stability of the dynamical system (\ref{sfrw12}) against linear perturbations, are the following,
\begin{equation}\label{stcondgg}
\frac{J_2}{J_1}>0,{\,}{\,}{\,}\frac{J_3}{J_1}>0\, ,
\end{equation}
where we introduced the variable $J_1$ which is equal to,
\begin{align}\label{st11}
J_1=288 g(N)^3 F''(G)\, ,
\end{align}
while the variable $J_2$ stands for,
\begin{align}\label{st12}
J_2=432 g(N)^{2 }\Big{(}(2 g(N)+g'(N)) F''(G)+8 g(N) \Big{(}g'(N)^2+g(N) (4 g'(N)+g''(N))\Big{)} F''(G)\Big{)}\, ,
\end{align}
and in addition, the parameter $J_3$ is equal to,
\begin{align}\label{st13}
& J_3=6 \Big{(}1+24 g(N)\Big{(}-8 g(N)^2+3 g'(N)^2+6 g(N) (3 g'(N)+g''(N))\Big{)}F''(G)
\\ \notag & +24 g(N) (4 g(N)+g'(N)) \Big{(}g'(N)^2+g(N) (4 g(N)+g''(N))\Big{)}F''(G)\Big{)}\, .
\end{align}
Having these at hand, and also the stability conditions, let us investigate whether the solution (\ref{fgsmalglimit}) is stable towards linear perturbations. The $F(G)$ gravity of Eq. (\ref{fgsmalglimit}) can be further simplified, since $\alpha$ satisfies the condition $1<\alpha<2$, so the most dominant term near the bouncing point (or equivalently at the small $G$ limit as we showed in the previous section), is the second term in Eq. (\ref{fgsmalglimit}) and hence the $F(G)$ function can be approximated by,
\begin{equation}\label{formfinalfgf}
F(G)\simeq B G^{\frac{\alpha }{-1+3 \alpha }}\, .
\end{equation}
By using the latter form of the $F(G)$ gravity, the variables $J_1$, $J_2$ and $J_3$ can easily be computed and we give their detailed form in the Appendix B. Note that the function $g(N)$ is expressed in terms of the $e$-folding number $N$ which is equal to $N=\ln a$, since we have set $a_0=1$, so for the Hubble rate of Eq. (\ref{hubratepresentpaper}), the function $g(N)$ reads,
\begin{equation}\label{gnfunctioform}
g(N)=\frac{\beta ^2 N^{\gamma }}{f_0}\, .
\end{equation}
where $\gamma=\frac{2\epsilon+1}{1+\epsilon}$. So finally, the stability conditions for the $F(G)$ gravity of Eq. (\ref{formfinalfgf}) read,
\begin{align}\label{st17}
\frac{J_2}{J_1}=\frac{3 \left(f_0^2 N (2 N+\gamma )+8 N^{2 \gamma } \beta ^4 \gamma  (-1+4 N+2 \gamma )\right)}{2 f_0^2 N^2}\, 
\end{align}
and moreover $J_3/J_1$ is equal to,
\begin{align}\label{st18}
&\frac{J_3}{J_1}=-\frac{1}{B f_0 \alpha  (-1+2 \alpha )}3^{1+\frac{\alpha }{1-3 \alpha }} 4^{\frac{\alpha }{1-3 \alpha }} N^{-2+\gamma } (1-3 \alpha )^2 \beta ^2 (\gamma )^2 \left(\frac{N^{-1+2 \gamma } \beta ^4 (\gamma )}{f_0^2}\right)^{\frac{\alpha }{1-3 \alpha }}\times \\ \notag &
\Big{(}1+\frac{2^{\frac{1-\alpha }{-1+3 \alpha }} 3^{\frac{1-2 \alpha }{-1+3 \alpha }} B f_0 N^{-\gamma } \alpha  (-1+2 \alpha ) \left(\frac{N^{-1+2 \gamma } \beta ^4 (2 N+\gamma )}{f_0^2}\right)^{\frac{\alpha }{-1+3 \alpha }} \left(8 N^2-18 N \gamma +3 (2-3 \gamma ) \gamma \right)}{(1-3 \alpha )^2 \beta ^2 (2 N+\gamma )^2}\\ \notag &
-\frac{2^{\frac{1-\alpha }{-1+3 \alpha }} 3^{\frac{1-2 \alpha }{-1+3 \alpha }} B \alpha  (-1+2 \alpha ) \left(\frac{N^{-1+2 \gamma } \beta ^4 (\gamma )}{f_0^2}\right)^{\frac{\alpha }{-1+3 \alpha }} (\gamma ) (\gamma  (-1+2 \gamma ))}{N (1-3 \alpha )^2 (\gamma )^2}\Big{)}\, .
\end{align}
Since we are interested in the behavior of $J_2/J_1$ and $J_3/J_1$ near the bouncing point, we should what the limit $t\rightarrow t_s$ means in terms of the $e$-folding number $N$. Actually, since when $t\rightarrow t_s$, the scale factor tends to unity, then, in the limit $t\rightarrow t_s$, the $e$-folding number tends to zero. Therefore, we shall find how $J_2/J_1$ and $J_3/J_1$ behave for $N\rightarrow 0$ and taking the limit of the expressions appearing in Eq. (\ref{st17}) and (\ref{st18}), we obtain,
\begin{align}\label{limitingcasesn0}
& \frac{J_2}{J_1}=3+\frac{3 \gamma }{2 N}\, , \\ \notag &
\frac{J_3}{J_1}=-\mathcal{A}N^{-2+\gamma +\frac{\alpha -2 \alpha  \gamma }{-1+3 \alpha }}\, ,
\end{align}
where the parameter $\mathcal{A}$ is positive, and can be found in the Appendix B. As it is obvious from Eq. (\ref{limitingcasesn0}), the parameter $J_3/J_1$ is negative and therefore the system is unstable, as we anticipated. This means that the solution (\ref{formfinalfgf}) is not a final attractor, and therefore it can describe the system near the bouncing point, but for a small time, since the system continues it's evolution, after that point. What now remains is to study the evolution of scalar perturbations of the cosmological evolution, with emphasis on the description near the Type IV singularity. This issue is addressed in the next section.

\section{Evolution of Scalar Perturbations}

Having the qualitative behavior of the $F(G)$ gravity that generates the singular bounce, it is worth examining how the scalar perturbations evolve, assuming the flat FRW background of Eq. (\ref{metricformfrwhjkh}). So we consider scalar linear perturbations of the flat FRW background of Eq. (\ref{metricformfrwhjkh}), of the form,
\begin{equation}\label{scalarpertrbubations}
\mathrm{d}s^2=-(1+\psi)\mathrm{d}t^2-2a(t)\partial_i\beta\mathrm{d}t\mathrm{d}x^i+a(t)^2\left( \delta_{ij}+2\phi\delta_{ij}+2\partial_i\partial_j\gamma \right)\mathrm{d}x^i\mathrm{d}x^j\, ,
\end{equation}
with $\psi$, $\phi$, $\gamma$ and $\beta$ being the smooth scalar perturbations. For the perturbation study we follow the approach and master equation given in Ref. \cite{felice}, but we specify everything for the $F(G)$ case, which is a special case of $F(R,G)$  gravity studied in \cite{felice}. Perturbations are usually analyzed in terms of gauge invariant quantities, for convenience reasons, therefore we shall be interested in the following gauge invariant quantity (the corresponding comoving curvature perturbation), the evolution of which we study in this section,
\begin{equation}\label{confedf}
\Phi=\phi-\frac{H\delta \xi}{\dot{\xi}}\, ,
\end{equation}
with $\xi=\frac{\mathrm{d}F}{\mathrm{d}G}$. In the $F(G)$ gravity case, the scalar modes propagating, contain no $k^4$ terms, so no superluminal propagation occurs, and only the usual $k^2$ terms appear \cite{felice}. The perturbation equation that governs the scalar modes in $F(G)$ gravity is the following,
\begin{equation}\label{perteqnmain}
\frac{1}{a(t)^3Q(t)}\frac{\mathrm{d}}{\mathrm{d}t}\left(a(t)^3Q(t)\dot{\Phi}\right)+B_1(t)\frac{k^2}{a(t)^2}\Phi=0\, ,
\end{equation}
where we can see the above equation has the usual form for scalar perturbations, in which $k^2$ terms dominate in the evolution. It is conceivable that the speed of propagation is determined by the term $B_1(t)$, which for $F(G)$ theories of gravity is defined to be,
\begin{equation}\label{b1}
B_1(t)=1+\frac{2\dot{H}}{H^2}\, .
\end{equation}
Moreover, the term $Q(t)$ appearing in Eq. (\ref{perteqnmain}), for the $F(G)$ case is equal to,
\begin{equation}\label{gfgdhbhhyhjs}
Q(t)=\frac{6\left( \frac{\mathrm{d}^2F}{\mathrm{d}G^2}\right )^2\dot{G}^2\left(1+4F''(G)\dot{G}H\right)}{\left(1+6HF''(G)\dot{G}\right)^2}\, ,
\end{equation}
where the prime denotes differentiation with respect to $G$, while the dot as usual denotes differentiation with respect to the cosmic time. Note additionally that we used the fact that $\dot{\xi}=\frac{\mathrm{d}F^2}{\mathrm{d}G^2}\dot{G}$.

It is conceivable that finding an analytic solution of Eq. (\ref{perteqnmain}) is rather difficult, so either a numerical study or an approximate solution is required. We shall choose the latter approach and seek for an approximate solution near the bouncing point. Before continue, we rewrite the differential equation (\ref{perteqnmain}) as follows,
\begin{equation}\label{difnew}
a(t)^3Q(t)\ddot{\Phi}+\left(3a(t)^2\dot{a}Q(t)+a(t)^3\dot{Q}(t)\right)\dot{\Phi}+B_1(t)Q(t)a(t)k^2\Phi=0\, .
\end{equation} 
After some tedious calculations, by using the resulting $F(G)$ gravity near the bouncing point, namely the one appearing in Eq. (\ref{fgsmalglimit}), and by keeping the most dominant terms near the bouncing point, the differential equation that governs the evolution of perturbations near $t=t_s$ reads,
\begin{equation}\label{simplifiedeqn1}
(t-t_s)^{1+\alpha } \Omega_4\ddot{\Phi}-(t-t_s)^{\alpha }\Omega_2\dot{\Phi}+\Omega_1\Phi=0\, ,
\end{equation}
where the parameters $\Omega_i$ ($i=1,2,4$) are given in the Appendix D. Note that we omitted a term $\sim \Omega_3 \dot{\Phi}(t-t_s)^{2 \alpha }$, which is subdominant compared to the term $(t-t_s)^{\alpha }\Omega_2\dot{\Phi}$. The parameter $\Omega_3$ can also be found in the Appendix D. Then, by solving the differential equation (\ref{difnew}) we obtain the following analytic solution which describes the evolution of scalar perturbations near the singular point $t=t_s$,
\begin{equation}\label{solutionevolution}
\Phi(t)= \Delta_1 x^{\frac{\mu }{2 (-1+\alpha )}}J_{\mu}(\zeta  x^{\frac{1-\alpha }{2}})+\Delta_2 x^{\frac{\mu }{2 (-1+\alpha )}}J_{-\mu}(\zeta  x^{\frac{1-\alpha }{2}})\, ,
\end{equation}
with $x=t-t_s$ and the constants $\mu$ and $\zeta$ depend on the parameters $\Omega_i$ and their full detailed form appears in the Appendix D. Also the function $J_{\mu}(y)$ is the Bessel function of the first kind and in addition, the parameters $\Delta_i$, $i=1,2$, are given below, 
\begin{align}\label{app1}
& \Delta_1=\left(-1+\frac{1}{\alpha }\right)^{\mu } \alpha ^{\mu } \Omega_1^{-\frac{\mu }{2}} \Omega_4^{\frac{\mu }{2}}\text{  }C_3 \Gamma\left[\frac{\alpha }{-1+\alpha }+\frac{\Omega_2}{(-1+\alpha ) \Omega_4}\right],\\ \notag & \Delta_2=\left(-1+\frac{1}{\alpha }\right)^{3\mu } \alpha ^{\mu } \Omega_1^{-\frac{\mu }{2}} \Omega_4^{\frac{\mu }{2}}C_4 \Gamma\left[\frac{\Omega_2}{\Omega_4-\alpha  \Omega_4}+\frac{2 \Omega_4}{\Omega_4-\alpha  \Omega_4}-\frac{\alpha  \Omega_4}{\Omega_4-\alpha  \Omega_4}\right]\, .
\end{align}
Note that in the parameters $\Delta_i$, $i=1,2$, appear the constants $C_3$ and $C_4$, which are arbitrary constants of integration, which result after solving the differential equation of Eq. (\ref{difnew}), without any initial conditions. Therefore, the solution of Eq. (\ref{solutionevolution}) is a general solution, and the constants of integration of this general solution are the parameters $C_3$ and $C_4$ appearing in the parameters $\Delta_i$, $i=1,2$ of Eq. (\ref{app1}). Since we are considering the limit $x\rightarrow 0$, we can further approximate the solution, by using the limit of the Bessel function for small arguments, which is,
\begin{equation}\label{besselapprox}
J_{\mu}(y)\simeq \frac{y^{\mu }2^{-\mu }}{\Gamma[1+\mu ]}\, ,
\end{equation}
so the approximate evolution of the scalar perturbations (\ref{solutionevolution}) near the Type IV singularity behaves as follows,
\begin{equation}\label{approximatebehavior}
\Phi(t)\simeq \Delta_2 \frac{2^{-\mu } \zeta ^{\mu }}{\Gamma[1+\mu ]}x^{\frac{\left(2-2 \alpha +\alpha ^2\right) \mu }{2 (-1+\alpha )}}\, .
\end{equation}
So the resulting expression for the evolution of perturbations in the absence of matter fluids is a described by a power-law function of the variable $x=t-t_s$. It is worth to further investigate the power spectrum and check whether it is scale invariant. Note however that we already are within an approximation and therefore we should stress that our results should considered only approximate and also that the full $F(G)$ solution is needed in order to answer the problem in a consistent way. Nevertheless, near the bouncing point it is still interesting to find how the power spectrum behaves within the context of the classical $F(G)$ theory. This may indicate how the full quantum description of the bounce theory will remedy any problematic issues. 

We start from the gauge invariant variable $\Phi$ given in Eq. (\ref{confedf}), which as was shown in \cite{felice}, satisfies the following second order perturbed action,
\begin{equation}\label{secondoredperturbas}
\mathcal{S}_p=\int \mathrm{d}x^4a(t)^3Q_s\left( \frac{1}{2}\dot{\Phi}-\frac{1}{2}\frac{c_s^2}{a(t)^2}(\nabla \Phi )^2\right)\, ,
\end{equation}
where $Q_s=\frac{4}{\kappa^2}Q(t)$ and $Q(t)$ appears in Eq. (\ref{gfgdhbhhyhjs}). Following the standard approach in perturbation theory \cite{mukhanov1aa,mukhanov1a,mukhanov}, the power spectrum of curvature perturbations for the scalar field $\Phi$ is,
\begin{equation}\label{powerspecetrumfgr}
\mathcal{P}_R=\frac{4\pi k^3}{(2\pi)^3}|\Phi|_{k=aH}^2\, .
\end{equation}
 It is straightforward to show that the power spectrum is not scale-invariant, by simply looking the forms of $\Omega_i$, $\Delta_2$ and $\zeta$ from Appendix D. As it can be seen, the wavenumber $k$ is contained only in $\Omega_1$, $\Delta_2$ and in $\zeta$ implicitly via $\Omega_1$. These have the following functional dependence with respect to $k$,
\begin{equation}\label{snddfr}
\Omega_1\sim k^2,\,\,\,\zeta\sim \sqrt{\Omega_1},\,\,\, \Delta_2\sim \Omega_1^{-\frac{\mu}{2}}\, ,
\end{equation}
and since the power spectrum depends on the combination $\Delta\zeta^{\mu}$, it follows from Eq. (\ref{snddfr}) that,
\begin{equation}\label{powerspectrajb}
\mathcal{P}_R\sim k^3 \Big{|} C_4 (t-t_s)^{\frac{\left(2-2 \alpha +\alpha ^2\right) \mu }{2 (-1+\alpha )}}\Big{|}^2_{k=aH}\, .
\end{equation} 
However, we cannot conclude at this point if the spectrum is scale invariant or not, since the parameter $C_4$ which is the constant of integration appearing in $\Delta_2$ in Eq. (\ref{app1}), depends on $k$, and its exact form will depend on the initial conditions of the vacuum form of $\Phi (t)$. In addition, the the term $(t-t_s)^{\frac{\left(2-2 \alpha +\alpha ^2\right) \mu }{2 (-1+\alpha )}}$ also has a dependence on $k$, since the power spectrum is computed at $k=aH$, that is, at the horizon crossing. So let us calculate these in detail. 

Before we start it is worth recalling that we are working for cosmological times for which $t-t_s\rightarrow 0$, and hence, the conformal time $\tau$, defined as $\mathrm{d}\tau =a^{-1}(t)\mathrm{d}t$, is approximately equal to $t$, since for $t-t_s\rightarrow 0$, the scale factor appearing in Eq. (\ref{scalebounce}), behaves as $a\simeq 1$. Also, since $a\sim 1$, for $t-t_s\rightarrow 0$, then we have that $k\simeq H$ at the horizon crossing, and this means that,
\begin{equation}\label{explicitequation}
\beta (t-t_s)^{\alpha}\simeq k\, ,
\end{equation}
which by solving with respect to $(t-t_s)$, yields,
\begin{equation}\label{solvedwithrespecttotts}
t-t_s\simeq \left (\frac{k}{\beta}\right )^{\frac{1}{\alpha}}\, .
\end{equation}
Hence, Eq. (\ref{solvedwithrespecttotts}) shows how $t-t_s$ behaves as a function of the wavenumber $k$ near the horizon crossing. We now proceed to find the $k$-dependence of $C_4$. To do so, we shall introduce the very frequently used in the literature \cite{mukhanov1aa,mukhanov1a} canonical variable $u=z_s \Phi$, with $z_s=Q(t)a(t)$, and since $a(t)\simeq 1$ for $(t-t_s)\rightarrow 0$, we have that $z_s\simeq Q(t)$, and therefore,
\begin{equation}\label{safakebelieve}
u\sim \Phi Q(t)\, ,
\end{equation}
and recall that $Q(t)$ is defined in Eq. (\ref{gfgdhbhhyhjs}). In terms of $u$, the action of Eq. (\ref{secondoredperturbas}) becomes near the bounce,
\begin{equation}\label{actiaonerenearthebounce}
\mathcal{S}_u\simeq \int \mathrm{d}^3\mathrm{d}\tau \left[ \frac{u'}{2}-\frac{1}{2}(\nabla u)^2+\frac{z_s''}{z_s}u^2\right ]\, ,
\end{equation}
where the prime indicates differentiation with respect to the conformal time, which as we demonstrated earlier is approximately identical to the cosmological time $t$, for $(t-t_s)\rightarrow 0$. Also the action of Eq. (\ref{actiaonerenearthebounce}), is defined modulo a factor of $a^{-1}$, which is approximately equal to one near the time instance $t\simeq t_s$. The vacuum state of the canonical field $u$ is the Bunch-Davies quantum fluctuating vacuum \cite{mukhanov1a} at exactly $t=t_s$, hence $u\sim \frac{e^{-ik\tau}}{\sqrt{k}}$. Note that imaginary phase will disappear when we shall compute the norm of the comoving curvature $|\Phi (t=t_s)|^2$, so the relevant form of $u$ for the $k$ dependence issue of $C_4$ is, $u\sim \frac{1}{\sqrt{k}}$. Owing to Eq. (\ref{safakebelieve}), we obtain that,
\begin{equation}\label{imanrun}
\Phi(t=t_s)\sim C_4\sim \frac{1}{\sqrt{k}Q(t)}\, .
\end{equation}
By using the fact that $F(G)$ for $(t-t_s)\rightarrow 0$, is approximated by Eq. (\ref{gupro}), we get that the function $Q(t)$ is approximately equal to,
\begin{equation}\label{isssoripiametaksytrelas}
Q(t)\simeq \frac{2^{-5+\frac{6 \alpha }{-1+3 \alpha }} 3^{-1+\frac{2 \alpha }{-1+3 \alpha }} B^2 (t-t_s)^{-4 \alpha } (1-2 \alpha )^2 \left(\alpha  \beta ^3\right)^{\frac{2 \alpha }{-1+3 \alpha }}}{(1-3 \alpha )^4 \beta ^6}\, ,
\end{equation}
and since $(t-t_s)$ is given by Eq. (\ref{solvedwithrespecttotts}), we get,
\begin{equation}\label{newsrevisison3}
Q(t(k)) \simeq \mathcal{Z}_1 k^{\frac{-4 \alpha }{\alpha}}=\mathcal{Z}_1 k^{-4}\, ,
\end{equation}
 where $\mathcal{Z}_1$ stands for,
 \begin{equation}\label{dfjedjsuestrsureend}
\mathcal{Z}_1= \frac{2^{-5+\frac{6 \alpha }{-1+3 \alpha }} 3^{-1+\frac{2 \alpha }{-1+3 \alpha }} B^2 (1-2 \alpha )^2 \left(\alpha  \beta ^3\right)^{\frac{2 \alpha }{-1+3 \alpha }}}{(1-3 \alpha )^4 \beta ^6\beta^{1/{\alpha}}}\, .
\end{equation}
By substituting $Q(t)$ from Eq. (\ref{newsrevisison3}) in Eq. (\ref{imanrun}), we finally obtain that that $C_4$ behaves as,
\begin{equation}\label{equationforc4final}
C_4\simeq \frac{1}{\mathcal{Z}_1}\frac{1}{\sqrt{k}k^{-4}}\sim k^{\frac{7}{2}}\, .
\end{equation}
So combining Eqs. (\ref{equationforc4final}), (\ref{solvedwithrespecttotts}) and (\ref{powerspectrajb}), we finally get that the $k$-dependence of the power spectrum $\mathcal{P}_R$ is of the form,
\begin{equation}\label{powerspectrumfinal}
\mathcal{P}_R\sim k^{\frac{7}{2}+3+\frac{\left(2-2 \alpha +\alpha ^2\right) \mu }{2 (-1+\alpha )}}\, ,
\end{equation}
and hence we conclude that the power spectrum is not scale-invariant. We can use the approximate value for the power spectrum of primordial curvature perturbations given in Eq. (\ref{powerspectrumfinal}), in order to calculate the spectral index of primordial curvature perturbations $n_s$ as follows: Combining the expression for $\mu$ given in
Eq.~(\ref{app}), and also the values of $\Omega_2$ and $\Omega_4$ given in Eq.~(\ref{longago}), we get that $\mu$ becomes $\mu=11/(1-\alpha)$. Therefore, the spectral index of primordial curvature perturbations is equal to,
\begin{equation}
 n_s-1\equiv\frac{d\ln\mathcal{P}_{\mathcal{R}}}{d\ln
k}=\frac{7}{2}+3+\frac{2-2\alpha+\alpha^2}{2(\alpha-1)}\mu=1-\frac{11}{2(\alpha-
1)^2}~.
\end{equation}
Therefore, the spectral index can be in agreement with the latest observational constraints on $n_s$ \cite{planck}, for
two values of $\alpha$: one which satisfies $\alpha<-1$ and
one which satisfies $\alpha>1$. The $\alpha<-1$ case leads to a Big Rip singularity, while the $\alpha > 1$ case leads to a Type IV singularity, which is the case we studied in this paper. However, the value of $\alpha$ would need to be quite large, and so it
would be a large deformation of a symmetric bounce which we assumed in this paper, therefore this result does not appear to be
very physical.

But still we need to interpret correctly this result, since what we have at hand is that the classical approximation of the $F(G)$ theory that describes the singular bounce (\ref{hubratepresentpaper}) near the bouncing point, fails to produce a scale invariant spectrum. This could be a strong motivation to use a LQC corrected $F(G)$ gravity theory \cite{sergeiharooikonomou} and investigate whether the same picture persists with quantum corrections added. If the answer lies in the affirmative, then this could probably mean that this effect is a feature of the Type IV singularity, that needs to be further investigated. 

However, if we use more physical values of $\alpha$ in the range $1<\alpha<2$, this would yield a very red spectrum, and one can strongly doubt that quantum corrections, e.g.~from LQC corrected $F(G)$ gravity, could adjust this to a nearly scale-invariant spectrum. This argument is very rigid and solid and therefore should be investigated to explicitly demonstrate its validity, a task that we defer for a future work.

\subsection{Discussion}

Before we close this section, we need to discuss in detail the physical results we obtained in this section. As we demonstrated, the power spectrum of primordial curvature perturbations is not scale invariant, when this is evaluated for cosmic times near the bouncing point at $t=t_s$, which is also the point at which the Type IV singularity occurs. However, we need to discuss the physical implications of this result. To this end, let us recall some fundamental issues for the dynamics of perturbations and related issues, but in the context of bouncing cosmology. Recall that the problem of initial conditions in the standard Big Bang cosmology was due to the fact that the Universe appears to be nearly flat and homogeneous in large scales, which cannot have causally communicated in the past. Therefore, a successful bouncing cosmology should in some way provide an elegant solution to these problems. 

Before we see what happens in the bouncing cosmology we studied in this paper, it is worth to present what happens in inflationary theories, so that we can compare the inflationary picture to the bouncing cosmology picture. In the inflationary cosmology picture, the primordial curvature perturbations \footnote{We are referring to the comoving curvature of Eq. (\ref{confedf}), so these are fluctuating vacuum scalar perturbations} which are of interest at present time, during inflation where at subhorizon scales, with horizon referring, usually in the existing related literature, to the Hubble radius $1/a(t)H(t)$. Then these perturbations freeze once they exit the horizon, which happens when the contracting horizon becomes comparable to their wavelength. Subsequently, these freezed perturbations become classical superhorizon perturbations which re-enter the horizon as the horizon expands again during the Hubble evolution of the Universe, after the reheating. Eventually, the gravitational collapse of the freezed superhorizon perturbations leads to the formation of the large-scale structure of the Universe and the Cosmic Microwave Background anisotropies correspond to superhorizon modes which were initially subhorizon during
inflation, and freezed after the horizon crossing. Moreover, anisotropies are due to modes which have re-entered the horizon, but still, these modes have the same origin on subhorizon scales during inflation.

At this point a more concise presentation is needed so that we gain deeper insights of the full picture. The novel feature of the inflationary description, is that during inflation, the Hubble radius, to which we refer as horizon, actually shrinks dramatically. So at the initial singularity the horizon was very large, and the primordial perturbations where actually as subhorizon scales, since their comoving wavenumber was at subhorizon scales, that is,
\begin{equation}\label{subhorizonsscalaesk}
k\gg H(t)a(t)\, .
\end{equation} 
Note that in principle perturbations are created at all length scales, but the most relevant are those which their wavenumber is at subhorizon scales. Also note that we switched our description by using the wavenumber, but it is equivalent in using the wavelength, in which case Eq. (\ref{subhorizonsscalaesk}) would be $\lambda \ll \left( H(t)a(t) \right)^{-1}$. During the inflationary era, the Hubble radius (the horizon) shrinks, so at some point the cosmologically relevant perturbations of wavenumber $k$ satisfying Eq. (\ref{subhorizonsscalaesk}) exit the horizon and freeze, meaning that these become classical. So these become superhorizon perturbations, in which case, their wavenumber satisfies,
\begin{equation}\label{superhorizon1}
k\ll a(t)H(t)\, .
\end{equation}
Once the horizon crossing occurs, the comoving curvature perturbations corresponding to the wavenumber $k$ cease to be of quantum nature, and the corresponding quantum expectation value of the comoving curvature perturbation is practically the classical ensemble stochastic average of a classical stochastic field. This is what we meant by freezing of these modes. The conservation of the average value of the comoving curvature perturbation in superhorizon scales is what actually enables us to relate the predictions corresponding to the time that the horizon is crossed, which actually corresponds to high energies, to the observable quantities corresponding to the horizon re-entry of the modes after reheating, which in turn corresponds to low energies. Note that the era between the horizon exit and re-entry is an era which is not fully understood, even up to date. Also this issue may appear also in bouncing cosmology, since in some cases what is needed is a quantum bounce description followed by some other model, see for example \cite{lcdmcai}. We shall discuss this issue in more detail and more concretely later on in this section. Coming back to the inflationary picture, at present time we are able to compute the inflationary observable quantities, because the subhorizon wavelengths freezed out at horizon exit and evolved in a classical way until nearly the present time era, after reheating and after the horizon re-entering. Note that, in principle, the primordial perturbations may freeze out well before the inflationary era ends, so these correspond to an expanding quantum era. Hence the primordial curvature perturbations corresponding to the expanding quantum era may be directly related to the Cosmic Microwave Background observables and also to other present time observables, since the quantum fluctuations freeze after the horizon exit. 

\begin{figure}[h] \centering
\includegraphics[width=18pc]{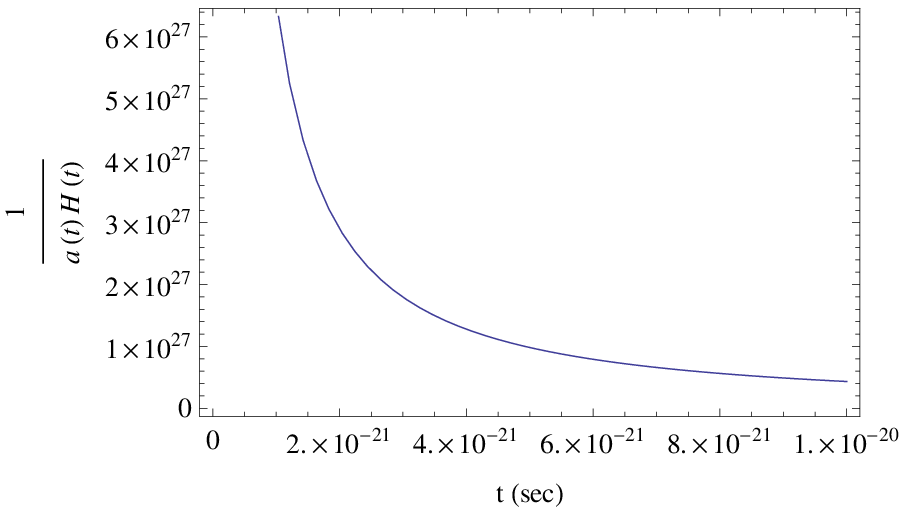}
\includegraphics[width=18pc]{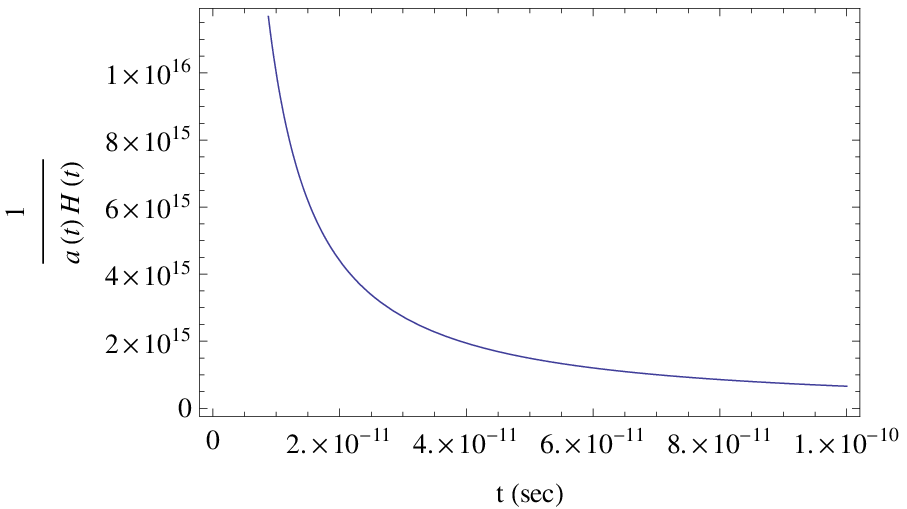}
\caption{The Hubble radius $r_H=\frac{1}{a(t)H(t)}$ as a function of time, for $t_s=10^{-35}$sec, $\alpha=13/11$, $\beta=0.001(\mathrm{sec})^{-\alpha-1}$. As it can be seen from the left and right plot, the Hubble radius $r_H$ fraction of corresponding to $t\simeq 10^{-20}$sec and $t\simeq 10^{-10}$sec is of the order $\frac{r_H(10^{-20})}{r_H(10^{-10})}\simeq 10^{12}$.}
\label{hubradpic}
\end{figure}

Let us now turn our focus to the singular bouncing cosmology of Eq. (\ref{hubratepresentpaper}). In this case, the Hubble radius as a function of the cosmological time is equal to,
\begin{equation}\label{hubbleradiusbounce}
r_H(t)=\frac{e^{-\frac{(t-t_s)^{1+\alpha } \beta }{1+\alpha }} (t-t_s)^{-\alpha }}{\beta }\, ,
\end{equation}
where for notational simplicity we denoted the Hubble radius as $r_H(t)=\frac{1}{a(t)H(t)}$. As it is obvious from Eq. (\ref{hubbleradiusbounce}), in the case of a Type IV singularity since $\alpha>1$, the Hubble radius at the bouncing point, which is also chosen to be the singularity point, the Hubble radius diverges due to the existence of the term $\sim (t-t_s)^{-\alpha }$. Hence all the cosmologically relevant modes are in subhorizon scales, since $k\gg H(t_s)a(t_s)=0$ at that point. Equivalently, at the singularity point, the cosmologically relevant modes are inside the Hubble radius which is infinite at that point, so the wavelength of these modes satisfies $\lambda \ll r_H$. Immediately after the bouncing point, the Hubble radius drops and starts to shrink as the time evolves. This can also be seen in Fig. \ref{hubradpic}, where we can see that the Hubble radius drops after the bouncing point, in a radical way. At the left plot, we plotted until $t\simeq 10^{-20}$sec and at the right plot until $t\simeq 10^{-10}$sec, and as it can be seen, the Hubble radius $r_H$ fraction corresponding to the two cases is of the order $\frac{r_H(10^{-20})}{r_H(10^{-10})}\simeq 10^{12}$.
Subsequently, as the Hubble radius shrinks, the cosmologically relevant modes will eventually exit the horizon, when the Hubble radius becomes of the order of their wavelength $\lambda\sim r_H$. But which modes can be cosmologically relevant in the singular bounce at hand? In principle, the time era near the bouncing point is governed by the quantum theory of gravity, so after the bouncing point we may still have the quantum era primordial modes. But to which cosmological times does the ``near the bounce'' expression refers to? Since we assume that $t_s\simeq 10^{-35}$sec, then near the bouncing point from a mathematical point of view, corresponds to cosmological times of the order $t\simeq 10^{-10}$sec, which can also be considered as being near the bounce, since what we assumed in the calculation of the spectrum of the primordial curvature perturbations is that $t-t_s\rightarrow 0$. Thereby, in inflationary terms these cosmic times correspond to times after the exit from the inflationary era, so the relevant for today modes have already exited the horizon well before $t\simeq 10^{-10}$sec. Hence for the singular bounce, the modes we took into account in the calculation of the spectrum of the primordial curvature perturbations, are actually the cosmologically relevant for today, quantum modes. Therefore, in the singular bounce case, the modes with wavelengths of the order of the horizon corresponding to cosmic times near the bouncing point, are the cosmologically relevant, since these can reveal the quantum era of primordial expansion. After these modes exit the Hubble radius, freeze out and the quantum expectation value of the comoving curvature perturbation is described by the classical ensemble stochastic average of a classical stochastic field. Practically, the conservation of the average value of the comoving curvature perturbation at superhorizon scales will eventually enable us to relate the horizon crossing predictions (high energy ones) to the late-time ones, which correspond to the horizon re-entry era (low energy). 

However we need to explicitly demonstrate that in the case of the singular bounce we studied in this paper, the comoving curvature perturbations are conserved after the modes exit the horizon. This is important, since it is not granted that the comoving curvature perturbations will be conserved, like for example in the inflationary cosmology. An example for which the comoving curvature perturbations grow after the modes exit the horizon is the matter bounce cosmology case \cite{mukhanov1aa}, so now we study the evolution of the comoving curvature perturbations in the context of the singular bounce. We are interested in cosmological times for which the Hubble rate and scale factor satisfy $k\ll a(t)H(t)$, so for times much more later than the horizon crossing. Note that this does not mean that $t$ corresponds to late-times, or $t\gg 1$, but the cosmic time is $t\gg t_H$, with $t_H$ the time at which the horizon crossing occurs. Since $k\ll a(t)H(t)$, this means that the last term in the differential equation of Eq. (\ref{perteqnmain}), can be neglected, so the differential equation becomes,
\begin{equation}\label{perteqnmainportoriko}
\frac{1}{a(t)^3Q(t)}\frac{\mathrm{d}}{\mathrm{d}t}\left(a(t)^3Q(t)\dot{\Phi}\right)=0\, ,
\end{equation} 
which can be easily solved and the solution is,
\begin{equation}\label{soldiffeqn}
\Phi (t)=\mathcal{C}_1+\mathcal{C}_2\int \frac{1}{a(t)^3Q(t)}\mathrm{d}t\, ,
\end{equation}
with $Q(t)$ being defined in Eq. (\ref{gfgdhbhhyhjs}). So in order to see if the comoving curvature perturbations are conserved after the horizon crossing, we need to examine the behavior of the integral term in Eq. (\ref{soldiffeqn}). Obviously, the key point to determine the behavior of the comoving curvature perturbations, is to determine $Q(t)$, and therefore to find which $F(G)$ gravity describes the cosmological evolution at the cosmological times for which $k\ll a(t)H(t)$. In order to proceed we need to classify the problem in two subcases or scenarios more preferably, which we list below:
\begin{enumerate}
    \item In the first scenario, which we call Scenario I, the cosmic times for which the relation $k\ll a(t)H(t)$, satisfy $t\gg t_s$, and also $t\ll 1$, where $t_s$ the time at which the Type IV singularity occurs. This is the most plausible scenario, from a physical point of view, and it could be realized like this: Suppose that $t_s=10^{-35}$sec and the time long after the horizon crossing is at $t=10^{-15}$sec. Hence, in this case, the time $t$ is $10^{20}$ times larger than $t_s$ and also satisfies $t\ll 1$. Note that the time at which horizon crossing occurs, is somewhere in between $t_s$ and $t$, that is, $t_s<t_H<t$. 
    \item In the second scenario, which we call Scenario II, we have again $t\gg t_s$, but $t>1$. 
\end{enumerate}    
The Scenario I is much more likely to occur, since the time for which $k\ll a(t)H(t)$ holds true, is after the horizon crossing, which for inflationary cosmology is (possibly) of the order $10^{-30}$sec, so this is the most appealing case and we start our analysis with Scenario I, but for completeness we also deal with Scenario II later on.

So in the context of Scenario I, the cosmic time satisfies $t\gg t_s$, but still $t\ll 1$, and consequently, in this case $t-t_s\rightarrow 0$. Therefore, the $F(G)$ gravity that is responsible for the cosmological evolution at the time $t$, is still given by Eq. (\ref{gupro}), and therefore, the function $Q(t)$ can easily be calculated to yield,
\begin{equation}\label{qtfunctionscenarioi}
Q(t)\simeq \mathcal{Z}_2 t^{-4\alpha}\, ,
\end{equation}
 where in Eq. (\ref{qtfunctionscenarioi}) we kept the most dominant term, and also we used the fact that $t-t_s\simeq t\ll 1$. Therefore, since $a(t)\simeq e^{f_0 t^{\alpha+1}}$ for the singular bounce (see Eq. (\ref{scalebounce})), the term $\frac{1}{a(t)^3Q(t)}$, behaves as follows,
\begin{equation}\label{integraterm1}
\frac{1}{a(t)^3Q(t)}\sim \frac{t^{4\alpha}}{e^{f_0 t^{\alpha+1}}}\, .
\end{equation}
Clearly, the exponential dominates, so the integral term decays as $t$ increases, that is,
\begin{equation}\label{shotshottohea}
\int \frac{1}{a(t)^3Q(t)}\rightarrow 0\, ,
\end{equation}
and hence, the comoving curvature perturbation is approximately equal to,
\begin{equation}\label{scomobafter}
\Phi (t)=\mathcal{C}_1\, ,
\end{equation}
which means that the comoving curvature perturbation for the Scenario I is conserved after the horizon crossing.

Now we turn our focus on Scenario II, for which $t\gg t_s$ and $t>1$. As we already mentioned, this scenario probably corresponds to an era much more later than the bounce, and possibly much more later than the horizon crossing, so it is rather less physically appealing. Regardless, we shall study this scenario for completeness. It is conceivable, that the $F(G)$ gravity which describes the cosmological evolution is no longer given by the one appearing in Eq. (\ref{gupro}), since $(t-t_s)$ is not small anymore. Therefore we need to find the $F(G)$ gravity that describes the cosmological evolution. This is however a formidable task, since by employing the reconstruction method we used in the previous sections, we end up to the following differential equation,
\begin{equation}\label{unsolvable}
2t^{\alpha+1}\ddot{P}(t)-2t^{1+2\alpha}\alpha \beta^4\dot{P}(t)+1=0\, ,
\end{equation}
which is very difficult to solve analytically. So in order to proceed, we speculate about the possible behavior of the $F(G)$ gravity. If for example the $F(G)$ gravity has polynomial form, that is $F(G)\sim B \,G^{\gamma}$, then, since for large $t$, the Gauss-Bonnet invariant becomes approximately equal to $G\sim t^{4\alpha}$, the resulting expression for $Q(t)$ is,
\begin{equation}\label{resulting1}
Q(t)\simeq \mathcal{Z}_3t^{-2+8 \alpha +8 \alpha  \gamma }\, ,
\end{equation}
with $\mathcal{Z}_3$ being equal to,
\begin{equation}\label{z3exact}
\mathcal{Z}_3=2^{12+6 \gamma } 3^{2+2 \gamma } B^2 \alpha ^2 \beta ^{8+8 \gamma }\, .
\end{equation}
Therefore, the term $\sim 1/(a(t)^3Q(t))$, is equal to,
\begin{equation}\label{termaq}
\frac{1}{a(t)^3Q(t)}\sim \frac{1}{e^{f_0 t^{\alpha+1}}t^{-2+8 \alpha +8 \alpha  \gamma }}\, .
\end{equation}
Hence, even in the case that $\gamma$ is a large negative real number, the exponential in the expression appearing in Eq. (\ref{termaq}) dominates and thereby the integral in Eq. (\ref{soldiffeqn}) decays and becomes subdominant. Hence in this case too, the comoving curvature perturbations after the horizon crossing are conserved, since $\Phi(t)=\mathcal{C}_1$.

In the case that the $F(G)$ is such so that $Q(t)$ dominates over the exponential scale factor, then the integral in Eq. (\ref{soldiffeqn}) dominates the evolution, and possibly the comoving curvature perturbations might grow as the time passes. For example, if $F(G)\sim e^{-G^{\gamma}}$, then the function $Q(t)$ becomes approximately equal to,
\begin{equation}\label{approxforgagt}
Q(t)\sim e^{- \mathcal{A}_1 t^{4 \alpha \gamma }} t^{-2+8 \alpha }\, ,
\end{equation}
with $\mathcal{A}_1=2^{1+3 \gamma } 3^{\gamma }\beta ^{4 \gamma }$, and therefore the term $\sim 1/(a(t)^3Q(t))$ becomes approximately equal to,
\begin{equation}\label{termaqbill}
\frac{1}{a(t)^3Q(t)}\sim \frac{e^{ \mathcal{A}_1 t^{4 \alpha \gamma }}}{e^{f_0 t^{\alpha+1}}t^{-2+8 \alpha }}\, ,
\end{equation}
which clearly does not decay. So the integral in the expression (\ref{soldiffeqn}) dominates the evolution of the comoving curvature perturbations after the horizon crossing, and the perturbations grow as the time grows larger. Therefore, in the case of the Scenario II, the curvature perturbations depend strongly on the form of the $F(G)$ gravity. However, the Scenario II is rather unlikely to occur, since the requirement $t>1\,$sec means that the Universe is at the lepton epoch, which is much more later than the time that the horizon crossing occurred. At the time $t>1$, it is possible that the singular bounce does not describe the Universe anymore, since another description must be found to generate the Hubble radius expansion, since the Hubble radius decreases in the context of the singular bounce. It is highly likely that a scenario like the one used in \cite{lcdmcai}, will take place and describe the Universe's evolution. Work is in progress for realizing such a scenario, but it is worth analyzing this a bit more. 

The problem with the singular bouncing cosmology occurs, because there exists no mechanism to make the Hubble radius increase eventually, unless we assume that the singular bounce governs early-times and after some time instance, the evolution is described by another scale factor, as in for example in Ref. \cite{lcdmcai}, see also \cite{lcdmsergei} for an $F(R)$ description of the model of Ref. \cite{lcdmcai}.  In the case of the model studied in Ref. \cite{lcdmcai}, the quantum radiation era was followed by a perfect fluid evolution, see \cite{lcdmcai}. So in the singular bouncing case, it is necessary to find another scenario that will describe the evolution of the Universe after the quantum bouncing era, for which new scenario, the re-entering of the modes in the expanding horizon will be possible. However, the in-between the two horizons era, leaves a gap in our description, and as we noted, this also occurs in the standard inflationary cosmology. This task  exceeds the purposes of this paper, but we hope in a future work to provide a model with two or more phases, for which a successful cosmological description may be achieved. However, this will not be in the context of $F(G)$ gravity, since as we demonstrated the spectrum of primordial curvature perturbations is not scale invariant, and this is a rather discouraging feature, but perhaps the LQC-corrected $F(G)$ gravity may have more appealing features.

\section{Singular Bounce from Mimetic $F(G)$ Gravity}

As a final study, we shall investigate which $F(G)$ gravity can generate the singular bounce of Eq. (\ref{hubratepresentpaper}), but in the context of mimetic $F(G)$ gravity. For a detailed account on this issue, see \cite{sergeioikonomoumimetic}. The mimetic $F(G)$ gravity approach uses the same action as the one appearing in Eq. (\ref{actionfggeneral}), but the Jordan frame metric is parameterized as follows \cite{mukhanov1,mukhanov2,Golovnev,Golovnev1},
\begin{equation}\label{100}
g_{\mu\nu}=-\hat{g}^{\rho\sigma}\partial_{\rho}\phi
\partial_{\sigma}\phi .
\end{equation}
Upon varying the metric tensor, we get,
$$
\delta g_{\mu\nu}=\hat{g}^{\rho\tau}\delta\hat{g}_{\tau\omega}
\hat{g}^{\omega\sigma}\partial_{\rho}\phi \partial_{\sigma}\phi
\hat{g}_{\mu\nu} -\hat{g}^{\rho\sigma}\partial_{\rho}\phi
\partial_{\sigma}\phi \delta\hat{g}_{\mu\nu}-
$$
$$
-2\hat{g}^{\rho\sigma}\partial_{\rho}\phi
\partial_{\sigma}\delta\phi \hat{g}_{\mu\nu}.
$$
and upon variation of the Jordan frame action (\ref{actionfggeneral}), with respect to
the redefined metric $\hat{g}_{\mu\nu}$, and with respect to the mimetic
scalar $\phi$, we obtain the following equations of motion,
\begin{equation}\label{eom}
R_{\mu\nu}- \frac{1}{2}R g_{\mu\nu}+
\end{equation}
$$
+8\Big[R_{\mu\rho\nu\sigma}+R_{\rho\nu}g_{\sigma\mu}-R_{\rho\sigma}g_{\nu\mu}-R_{\mu\nu}g_{\sigma\rho}+R_{\mu\sigma}g_{\nu\rho}
+\frac{R}{2}\left(g_{\mu\nu}g_{\sigma\rho}-g_{\mu\sigma}g_{\nu\rho}\right)\Big]\nabla^{\rho}\nabla^{\sigma}F_{\cal
G}+ \left(F_{G}{G}-F({G})\right)g_{\mu\nu}+
$$
$$
+\partial_{\mu}\phi\partial_{\nu}\phi\left(-R+8\left(-R_{\rho\sigma}
+\frac{1}{2}R
g_{\rho\sigma}\right)\nabla^{\rho}\nabla^{\sigma}F_{\cal
G}+4(F_{G}{G}-F({G}))\right)=
$$
$$
=T_{\mu\nu}+\partial_{\mu}\phi\partial_{\nu}\phi T,
$$
with $F_{G}$ standing for $F_{G}=d F({G}) / d{ G}$. Moreover, upon variation of the action (\ref{actionfggeneral}) with respect to the mimetic scalar
field $\phi$, we get,
\begin{equation}\label{FG}
\nabla^{\mu}\left(\partial_{\mu}\phi\left(-R+8\left(-R_{\rho\sigma}+\frac{1}{2}
R g_{\rho\sigma}\right)\nabla^{\rho}\nabla^{\sigma}F_{\cal
G}+4(F_{G}{G}-F({G})\right)-T)\right)=0.
\end{equation}

Since the following relation holds true \cite{sergeioikonomoumimetic},
\begin{equation}\label{introeqns}
g^{\mu\nu}\partial_{\mu}\phi\partial_{\nu}\phi=-1
\end{equation}
and owing to the fact that the mimetic scalar $\phi$ depends only on the cosmic time, we get the constraint $\phi=t$. Hence the $(t,t)$ component of
the expression appearing in Eq. (\ref{eom}) becomes,
\begin{equation}\label{FRIED-1}
2\dot{H}+3H^{2}+16H(\dot{H}+H^{2})\frac{d F_{{\cal
G}}}{dt}+8H^{2}\frac{d^2 F_{{G}} }{dt^2} -(F_{{G}}{\cal
G}-F({G}))=-p.
\end{equation}
Upon integration of Eq. (\ref{FG}), we get,
\begin{equation}
-R+8\left(-R_{\rho\sigma}+\frac{1}{2}R
g_{\rho\sigma}\right)\nabla^{\rho}\nabla^{\sigma}F_{{\cal
G}}+4(F_{{G}}{G}-F({G}))+\rho-3p=-\frac{C}{a^3}
\end{equation}
which can be rewritten as follows,
\begin{equation}\label{FRIED-2}
\dot{H}+2H^{2}+4H^{2}\frac{d^{2}F_{{G}}}{dt^2}+
4H\left(2\dot{H}+3H^{2}\right)\frac{d F_{{\cal
G}}}{dt}+\frac{2}{3}(F_{{G}}{G}-F({G}))+
\frac{\rho}{6}-\frac{p}{2}=-\frac{C}{a^3}.
\end{equation}
where $C$ being an arbitrary constant.

The combination of Eqs. (\ref{FRIED-1}) and (\ref{FRIED-2}), results in,
\begin{equation}\label{fried3}
\dot{H}+4H^{2}\frac{d^{2}F_{{\cal
G}}}{dt^2}+4H(2\dot{H}-H^{2})\frac{dF_{{\cal
G}}}{dt}=-\frac{1}{2}\left(\rho+p\right)-\frac{C}{a^{3}}.
\end{equation}
For convenience, we introduce  the function $g(t)$, which is defined to be,
$$
g(t)=\frac{dF_{{G}}}{dt},
$$
and satisfies the following equation,
\begin{equation}\label{FRIED4}
4H^{2}\frac{dg(t)}{dt}+4H(2\dot{H}-H^2)g(t)=-\dot{H}-\frac{1}{2}\left(\rho+p\right)-\frac{C}{a^{3}}.
\end{equation}
Then, the general form of the solution of the differential equation (\ref{FRIED4}) is,
\begin{equation}
g(t)=g_{0}\left(\frac{H_{0}}{H}\right)^{2}\exp\left(\int_{{0}}^{t}Hdt\right)+\frac{1}{4H^{2}}\int_{0}^{t}dt_{1}B(t_{1})\exp\left(\int_{t_{1}}^{t}
H(\tau)d\tau\right).\label{genericbouncesol}
\end{equation}
where $B(t)=-\dot{H}-\frac{1}{2}\left(\rho+p\right)-\frac{C}{a^{3}}$ and in addition $g_{0}$ is an integration constant, while, $H_{0}=H({0})$. Having Eq. (\ref{genericbouncesol}), for a given cosmological evolution in terms of the Hubble rate, we can easily obtain the function $g(t)$. Consequently, the function $F(t)$ reads,
$$
F_{{G}}(t)=\int g(t) dt.
$$
Then by exploiting the expression (\ref{gausbonehub}), and solving with respect to $t$, we have the function $t=t(G)$. By substituting this
to $F_G(t)$, and by integrating with respect to $G$, we easily obtain the $F(G)$ function. Let us apply this method for the case of the singular bounce of Eq. (\ref{hubratepresentpaper}). By substituting the Hubble rate of Eq. (\ref{hubratepresentpaper}), in Eq. (\ref{genericbouncesol}), and by keeping the most dominant terms, we obtain,
\begin{equation}\label{fki}
g(t)=\frac{e^{f_0 (t-t_s)^{1+\alpha }} g_0 H_0^2 (t-t_s)^{-2 \alpha }}{f_0^2 (1+\alpha )^2}+\frac{(t-t_s)^{-2 \alpha } \left(-\frac{C t^2}{2}-\frac{p t^2}{4}-\frac{t^2 \rho }{4}\right)}{4 f_0^2 (1+\alpha )^2}\, .
\end{equation}
Integrating this with respect to $t$, we obtain the approximate form of the function $F_G(t)$ near the bouncing point, which is,
\begin{align}\label{gfgdv}
& F_G(t)=\frac{(t-t_s)^{1-2 \alpha } \left(-16 g_0 H_0^2  \left(3-5 \alpha +2 \alpha ^2\right)+ \left(t_s^2+t (t_s-2 t_s \alpha )+t^2 \left(1-3 \alpha +2 \alpha ^2\right)\right) (2 C+p+\rho )\right)}{16 f_0^2 (-1+\alpha ) (1+\alpha )^2 (-3+2 \alpha ) (-1+2 \alpha )}\\ \notag &
+\frac{f_0^{-\frac{3}{1+\alpha }} g_0 H_0^2 \Gamma\left[\frac{3}{1+\alpha }\right]}{(1+\alpha ) \left(2-5 \alpha +2 \alpha ^2\right)}\, .
\end{align}
Using (\ref{gausbonehub}), and also Eq. (\ref{onyfe}), we obtain the function $F_G(G)$, so upon integration with respect to $G$ we finally get,
\begin{align}\label{finalfgmimetic}
& F(G)\simeq -24 f_0^3 \alpha  (1+\alpha )^3 \left(a_7 G^{-\frac{3 \alpha }{1-3 \alpha }}\right)\\ \notag &
+\frac{-24 f_0^3 \alpha  (1+\alpha )^3\text{  }a_1}{a_6} G^{\frac{1+\alpha }{-1+3 \alpha }} \left(a_2+3+a_4 \left(t_s+24^{\frac{1}{1-3 \alpha }} G^{\frac{1}{-1+3 \alpha }} \left(-\frac{1}{f_0^3 \alpha  (1+\alpha )^3}\right)^{\frac{1}{-1+3 \alpha }}\right)\right)\\ \notag &
+\frac{-24 f_0^3 \alpha  (1+\alpha )^3 (2 C+p+\rho ) a_5 a_1 G^{\frac{1+\alpha }{-1+3 \alpha }}}{a_6} \left(t_s+24^{\frac{1}{1-3 \alpha }} G^{\frac{1}{-1+3 \alpha }} \left(-\frac{1}{f_0^3 \alpha  (1+\alpha )^3}\right)^{\frac{1}{-1+3 \alpha }}\right)^2\, ,
\end{align}
where the parameters $a_i$, $i=1,2,...7$ are given in Appendix C. Since we are considering the limit $t\rightarrow t_s$, which means $G\rightarrow 0 $ as we explained earlier, by keeping the most dominant term in Eq. (\ref{finalfgmimetic}), we get,
\begin{equation}\label{dfeslimitingcase}
F(G)=-\frac{24 f_0^3 \alpha  (1+\alpha )^3 (2 C+p+\rho )t_s^2 a_5 a_1 }{a_6}G^{\frac{1+\alpha }{-1+3 \alpha }}\, .
\end{equation}
It is conceivable that the resulting mimetic $F(G)$ gravity of Eq. (\ref{dfeslimitingcase}) is different for the vacuum $F(G)$ gravity of Eq. (\ref{fgsmalglimit}), but we need to further analyze this issue. In the context of the mimetic $F(G)$ gravity, some extra conformal degrees of freedom arise in the FRW equations of motion. Therefore, in the context of mimetic $F(G)$ gravity, we have a new reconstruction method in which we can choose the internal degrees of freedom and a specific $F(G)$ gravity so that some fixed cosmological evolution is generated. This is different from the ordinary vacuum $F(G)$ gravity case, since in this case no internal degrees of freedom are taken into account, so the $F(G)$ gravity that can generate the same Hubble rate as the mimetic $F(G)$ does, is in principle different from the resulting expression of the mimetic $F(G)$ gravity. Of course in both cases we are using approximations, so one should be cautious when dealing with both theories. Finally, let us note that in the mimetic $F(G)$ gravity, much more freedom is offered for successfully generating various cosmological scenarios, see for example \cite{Golovnev1}. This is because of the presence of these internal conformal degrees of freedom. It is then easy to reconstruct any cosmology by suitably adjusting these degrees of freedom, and in some approaches the potential and the Lagrange multiplier. It is questionable however if these results can be trusted, because the resulting picture is complicated. If simplicity is to be an important feature of a physical theory, then probably these theories are of mathematical importance only. However, concordance with observations is an appealing feature, and therefore these theories can be valuable from a physical point of view. Since this discussion should be addressed in more detail, we defer it to a future work. Regardless of the differences, in both cases, the $F(G)$ gravity that can realize the singular bounce of Eq. (\ref{hubratepresentpaper}), is described by a power-law function. But in our opinion, the vacuum $F(G)$ is conceptually a simpler theory, so from this aspect it offers a more appealing physical description of the singular bounce.

\section{Conclusions}

In this paper we studied a bounce cosmology with a Type IV singularity at the bouncing point, in the context of classical $F(G)$ gravity. Particularly, we investigated which classical pure (vacuum) $F(G)$ gravity can generate the Type IV singular bounce cosmology, emphasizing for cosmic times near the bouncing point. As we explicitly demonstrated, the resulting $F(G)$ gravity has the form $F(G)\sim C_2 G+B\, G^{\frac{\alpha }{-1+3 \alpha }}$, so it is a power-law modified gravity theory. Since this result holds true only near the singularity point, we discussed the possibility that this $F(G)$ gravity is the limiting case of some viable $F(G)$ gravity, in which case the full solution would also be interesting, since the late-time and early-time acceleration could be simultaneously described by the same theory. We also discussed the stability of the resulting $F(G)$ theory, from a dynamical point of view, examining if it can be the final attractor of the theory. As we anticipated, the answer to this question does not lie in the affirmative, and hence instability of the solution cannot be avoided. This feature is welcome, since the cosmological evolution does not stop at the bouncing point, and therefore the resulting $F(G)$ gravity was not anticipated to be a stable solution of the cosmological system. Moreover, we investigated how the scalar cosmological perturbations of the background flat FRW metric, behave near the bouncing point, and we explicitly calculated the spectrum of primordial curvature perturbations. As we showed, the spectrum is not scale invariant and as we claimed in the main text, this result should be further investigated. This is due to the fact that we cannot be sure if it is a universal feature of the theory that owes its existence in the Type IV singularity, or it is an artifact of the approximations we made to obtain the resulting $F(G)$ gravity. The latter seems more plausible, however this feature has to be thoroughly addressed. Another important point that we need to stress, with regards to the non-scale invariance of the spectrum of primordial curvature perturbations, is that since we are studying a classical theory near the bouncing point, it might be possible that at these cosmic time scales, quantum effects take place. So effectively the lack of scale invariance in the power spectrum might be an effect of the classical approach to the problem, so the same problem should be addressed in the context of Loop Quantum Cosmology \cite{LQC}, and particularly in the context of $F(G)$ LQC, which was developed in \cite{sergeiharooikonomou}. We also studied which mimetic $F(G)$ gravity can describe the singular bounce near the bouncing point, by adopting the formalism of \cite{sergeioikonomoumimetic} and the resulting $F(G)$ gravity has a power-law functional form.

Finally, with regards to the classical $F(G)$ gravity approach, since the $F(G)$ gravity is a special case of the most general class of $F(R,G)$ theory \cite{fg5,fg6,felice,frg,cappofrg}, the same problem we investigated in this problem should be addressed in the context of $F(R,G)$ theory. Actually, this problem should also be compared with the $F(R,G)$ gravity inflation properties, as was done in \cite{cappofrg}, but this time by using a Type IV singular bounce. In addition, a compelling task is to include matter fluids in the theory and investigate how the physical picture is affected by the presence of matter. Moreover, as was demonstrated in Ref \cite{noo4} that a Type IV may play a crucial role in the graceful exit from inflation, but the study was focused on scalar field models. It is worth to examine the effect of the Type IV singularity on Jordan frame $F(G)$ theories but also in $F(R)$ and $F(T)$ theories. For a related work with regards to the latter, see \cite{nashed}. We hope to materialize these projects in a future work.

\section*{Acknowledgments}

This work is supported by Min. of Education and Science of Russia (V.K.O).

\section*{Appendix A: Analytic Form of $Q(t)$ and of $A,B$}

In this Appendix we quote the exact form of the function $Q(t)$ and of $A,B$ appearing in Eqs. (\ref{dfesolu}) and (\ref{actaulfg}). Particularly, the function $Q(t)$ reads,
\begin{align}\label{analyticqt}
& Q(t)=-2 (t-t_s)^{-1+2 \alpha } \alpha  \beta ^2-72 (t-t_s)^{-1+3 \alpha } \alpha  \beta ^3\times \\ \notag &
\Big{(}-\frac{(t-t_s)^{1-2 \alpha } \left((t-t_s)^{-1+\alpha } \alpha -2 (t-t_s)^{-1+\alpha } \alpha ^2\right)}{2 (-1+\alpha ) (-1+2 \alpha ) \beta }\\ \notag &-\frac{(t-t_s)^{-2 \alpha } (1-2 \alpha ) \left((t-t_s)^{\alpha } -2 (t-t_s)^{\alpha } \alpha -2 C_1 \beta +2 C_1 \alpha  \beta \right)}{2 (-1+\alpha ) (-1+2 \alpha ) \beta }\Big{)}\\ \notag &
24 (t-t_s)^{3 \alpha } \beta ^3\Big{(} -\frac{(t-t_s)^{-2 \alpha } (1-2 \alpha ) \left((t-t_s)^{-1+\alpha } \alpha -2 (t-t_s)^{-1+\alpha } \alpha ^2\right)}{(-1+\alpha ) (-1+2 \alpha ) \beta }\\ \notag &
-\frac{(t-t_s)^{1-2 \alpha } \left((t-t_s)^{-2+\alpha } (-1+\alpha ) \alpha -2 (t-t_s)^{-2+\alpha } (-1+\alpha ) \alpha ^2\right)}{2 (-1+\alpha ) (-1+2 \alpha ) \beta }\\ \notag & +\frac{(t-t_s)^{-1-2 \alpha } (1-2 \alpha ) \alpha  \left((t-t_s)^{\alpha }-2 (t-t_s)^{\alpha } \alpha -2 C_1 \beta +2 C_1 \alpha  \beta \right)}{(-1+\alpha ) (-1+2 \alpha ) \beta }\Big{)}\, .
\end{align}
Also, the exact analytic form of the coefficients $A$ and $B$ appearing in Eq. (\ref{actaulfg}) is the following,
\begin{align}\label{dgdgdg}
& A=11\ 24^{\frac{2 \alpha }{1-3 \alpha }} \beta ^2 \left(\alpha  \beta ^3\right)^{\frac{2 \alpha }{1-3 \alpha }}\\ \notag &
B= \frac{\left(24^{\frac{1-2 \alpha }{1-3 \alpha }} C_1-24^{\frac{1-2 \alpha }{1-3 \alpha }} C_1 \alpha -\frac{2^{-1+\frac{3 (1-2 \alpha )}{1-3 \alpha }} 3^{\frac{1-2 \alpha }{1-3 \alpha }}}{\beta }+\frac{24^{\frac{1-2 \alpha }{1-3 \alpha }} \alpha }{\beta }\right) \left(\alpha  \beta ^3\right)^{\frac{1-2 \alpha }{1-3 \alpha }}}{1-3 \alpha +2 \alpha ^2}\\ \notag & -24^{1+\frac{\alpha }{1-3 \alpha }} C_1 G^{\frac{\alpha }{-1+3 \alpha }} \beta ^3 \left(\alpha  \beta ^3\right)^{\frac{\alpha }{1-3 \alpha }}
\end{align}

\section*{Appendix B: Exact Form of the Parameters $J_1,J_2,J_3$ and of $\mathcal{A}$}

Here we quote the exact form of the parameters $J_1,J_2,J_3$ and of $\mathcal{A}$. Particularly, the parameter $J_1$ is,
\begin{align}\label{st14}
&J_1=\frac{2^{1+\frac{2 \alpha }{-1+3 \alpha }} 3^{\frac{\alpha }{-1+3 \alpha }} B N^{\gamma } \alpha ^2 \beta ^2 \left(\frac{N^{\gamma } \beta ^2 \left(\frac{2 N^{\gamma } \beta ^2}{f_0}+\frac{N^{-1+\gamma } \beta ^2 \gamma }{f_0}\right)}{f_0}\right)^{\frac{\alpha }{-1+3 \alpha }}}{f_0 (-1+3 \alpha )^2 \left(\frac{2 N^{\gamma } \beta ^2}{f_0}+\frac{N^{-1+\gamma } \beta ^2 \gamma }{f_0}\right)^2}\\ \notag & -\frac{2^{1+\frac{2 \alpha }{-1+3 \alpha }} 3^{\frac{\alpha }{-1+3 \alpha }} B N^{\gamma } \alpha  \beta ^2 \left(\frac{N^{\gamma } \beta ^2 \left(\frac{2 N^{\gamma } \beta ^2}{f_0}+\frac{N^{-1+\gamma } \beta ^2 \gamma }{f_0}\right)}{f_0}\right)^{\frac{\alpha }{-1+3 \alpha }}}{f_0 (-1+3 \alpha ) \left(\frac{2 N^{\gamma } \beta ^2}{f_0}+\frac{N^{-1+\gamma } \beta ^2 \gamma }{f_0}\right)^2}\, ,
\end{align}
while $J_2$ is,
\begin{align}\label{st15}
& J_2=\frac{1}{f_0^2}432 N^{2 \gamma } \beta ^4\left(\frac{N^{\gamma } \beta ^2 \left(\frac{2 N^{\gamma } \beta ^2}{f_0}+\frac{N^{-1+\gamma } \beta ^2 \gamma }{f_0}\right)}{f_0}\right)^{-2+\frac{\alpha }{-1+3 \alpha }}\times \\ \notag &
\Big{(}\frac{12^{-2+\frac{\alpha }{-1+3 \alpha }} B \alpha  \left(-1+\frac{\alpha }{-1+3 \alpha }\right) \left(\frac{2 N^{\gamma } \beta ^2}{f_0}+\frac{N^{-1+\gamma } \beta ^2 \gamma }{f_0}\right) }{-1+3 \alpha }\\ \notag &
+\frac{2^{-1+\frac{2 \alpha }{-1+3 \alpha }} 3^{-2+\frac{\alpha }{-1+3 \alpha }} B N^{\gamma } \alpha  \left(-1+\frac{\alpha }{-1+3 \alpha }\right) \beta ^2\text{  }\left(\frac{N^{-2+2 \gamma } \beta ^4 \gamma ^2}{f_0^2}+\frac{N^{\gamma } \beta ^2 \left(\frac{4 N^{-1+\gamma } \beta ^2 \gamma }{f_0}+\frac{N^{-2+\gamma } \beta ^2 (-1+\gamma ) \gamma }{f_0}\right)}{f_0}\right)}{f_0 (-1+3 \alpha )}
\Big{)}\, ,
\end{align}
and finally $J_3$ is,
\begin{align}\label{st16}
& J_3=62^{-1+\frac{2 \alpha }{-1+3 \alpha }} 3^{-1+\frac{\alpha }{-1+3 \alpha }} B N^{\gamma } \alpha  \left(-1+\frac{\alpha }{-1+3 \alpha }\right) \beta ^2 \left(\frac{N^{\gamma } \beta ^2 \left(\frac{2 N^{\gamma } \beta ^2}{f_0}+\frac{N^{-1+\gamma } \beta ^2 \gamma }{f_0}\right)}{f_0}\right)^{-2+\frac{\alpha }{-1+3 \alpha }}\\ \notag &
\Big{(}1+\frac{ \left(\frac{4 N^{\gamma } \beta ^2}{f_0}+\frac{N^{-1+\gamma } \beta ^2 \gamma }{f_0}\right) \left(\frac{N^{-2+2 \gamma } \beta ^4 \gamma ^2}{f_0^2}+\frac{N^{\gamma } \beta ^2 \left(\frac{4 N^{\gamma } \beta ^2}{f_0}+\frac{N^{-2+\gamma } \beta ^2 (-1+\gamma ) \gamma }{f_0}\right)}{f_0}\right)}{f_0 (-1+3 \alpha )}\\ \notag &
+\frac{ \left(-\frac{8 N^{2 \gamma } \beta ^4}{f_0^2}+\frac{3 N^{-2+2 \gamma } \beta ^4 \gamma ^2}{f_0^2}+\frac{6 N^{\gamma } \beta ^2 \left(\frac{3 N^{-1+\gamma } \beta ^2 \gamma }{f_0}+\frac{N^{-2+\gamma } \beta ^2 (-1+\gamma ) \gamma }{f_0}\right)}{f_0}\right)}{f_0 (-1+3 \alpha )}\Big{)}\, .
\end{align}
Also the parameter $\mathcal{A}$ which appears in Eq.  (\ref{limitingcasesn0}), is equal to,
\begin{equation}\label{gfredhotlady}
\mathcal{A}=\frac{3^{1+\frac{\alpha }{1-3 \alpha }} 4^{\frac{\alpha }{1-3 \alpha }} f_0 (1-3 \alpha )^2 \gamma  \left(\frac{\beta ^4 \gamma }{f_0^2}\right)^{1+\frac{\alpha }{1-3 \alpha }}}{B \alpha  (-1+2 \alpha ) \beta ^2}\, .
\end{equation}

\section*{Appendix C: Detailed Form of the Parameters $a_1,a_2,a_3,a_4,a_5,a_6,a_7$}

The detailed form of the parameters $a_i$, $i=1,2,...7$ that appear in Eq. (\ref{finalfgmimetic}) are,
\begin{align}\label{parametersai}
& a_1=\left(24^{\frac{1}{1-3 \alpha }} \left(-\frac{1}{f_0^3 \alpha  (1+\alpha )^3}\right)^{\frac{1}{-1+3 \alpha }}\right)^{1+\alpha },\\ \notag &
a_2=-16 g_0 H_0^2 \left(18-15 \alpha -10 \alpha ^2+5 \alpha ^3+2 \alpha ^4\right)\\ \notag &
a_3=t_s^2 \left(11-6 \alpha +3 \alpha ^2\right), \\ \notag &
a_4=t_s \left(5-6 \alpha -9 \alpha ^2+2 \alpha ^3\right)\\ \notag &
a_5=\left(2-3 \alpha -4 \alpha ^2+3 \alpha ^3+2 \alpha ^4\right),\\ \notag & a_6=16 f_0^2 (1+\alpha )^3 (-1+2 \alpha ) \left(-2+\alpha +\alpha ^2\right) \left(-9+3 \alpha +2 \alpha ^2\right)\\ \notag &
a_7=\frac{f_0^{-\frac{3}{1+\alpha }} g_0 H_0^2 \left(24^{\frac{1}{1-3 \alpha }} \left(-\frac{1}{f_0^3 \alpha  (1+\alpha )^3}\right)^{-\frac{1}{1-3 \alpha }}\right)^{3 \alpha }\Gamma\left[\frac{3}{1+\alpha }\right] }{6 \alpha -9 \alpha ^2-9 \alpha ^3+6 \alpha ^4}\, .
\end{align}

\section*{Appendix D: The Parameters $\Omega_1$, $\Omega_2$, $\Omega_3$, $\Omega_4$, $\mu$ and $\zeta$}

Here we quote the detailed form of the parameters $\Omega_1$, $\Omega_2$, $\Omega_3$, $\Omega_4$ and $\Delta_i$, $i=1,2$, $\mu$ and $\zeta$. Particularly, the parameters $\Omega_1$, $\Omega_2$, $\Omega_3$, $\Omega_4$ appearing in Eq. (\ref{simplifiedeqn1}) are equal to,
\begin{align}\label{longago}
& \Omega_1=\frac{2^{-4+\frac{6 \alpha }{-1+3 \alpha }} 3^{-1+\frac{2 \alpha }{-1+3 \alpha }} B^2 k^2\text{  }(1-2 \alpha )^2 \alpha  \left(\alpha  \beta ^3\right)^{\frac{2 \alpha }{-1+3 \alpha }} }{(1-3 \alpha )^4 \beta ^7}, \\ \notag &
\Omega_2= -\frac{2^{-3+\frac{6 \alpha }{-1+3 \alpha }} 3^{-1+\frac{2 \alpha }{-1+3 \alpha }} B^2\text{  }(1-2 \alpha )^2 \alpha  \left(\alpha  \beta ^3\right)^{\frac{2 \alpha }{-1+3 \alpha }}}{(1-3 \alpha )^4 \beta ^6}, \\ \notag &
\Omega_3=\frac{2^{-5+\frac{6 \alpha }{-1+3 \alpha }} 3^{\frac{2 \alpha }{-1+3 \alpha }} B^2\text{  }(1-2 \alpha )^2 \alpha  \left(\alpha  \beta ^3\right)^{\frac{2 \alpha }{-1+3 \alpha }}}{(1-3 \alpha )^4 (1+\alpha ) \beta ^5}, \\ \notag &
\Omega_4=\frac{2^{-5+\frac{6 \alpha }{-1+3 \alpha }} 3^{-1+\frac{2 \alpha }{-1+3 \alpha }} B^2 (1-2 \alpha )^2 \left(\alpha  \beta ^3\right)^{\frac{2 \alpha }{-1+3 \alpha }} }{(1-3 \alpha )^4 \beta ^6}\, .
\end{align}
In addition, the parameters $\mu$ and $\zeta$ appearing in Eq. (\ref{solutionevolution}) are equal to,
\begin{align}\label{app}
& \mu =\frac{\Omega_2+\Omega_4}{(-1+\alpha ) \Omega_4},
\\ \notag & \zeta =\frac{2\text{  }\sqrt{\Omega_1}}{\left(-1+\frac{1}{\alpha }\right) \alpha  \sqrt{\Omega_4}}
\end{align}

\end{document}